\documentclass[longbibliography,aps,prl,twocolumn]{revtex4-1}

\usepackage{graphicx} 
\usepackage[english]{babel}
\usepackage[T1]{fontenc}
\usepackage[usenames,dvipsnames]{xcolor}
\usepackage{setspace}
\usepackage[compact]{titlesec}
\usepackage{hyperref}
\usepackage{tikz}
\usepackage{amsmath}
\usepackage{amssymb}
\usepackage{mathtools}
\usepackage{appendix}
\begin{document}

\title{Rationalizing Euclidean Assemblies of Hard Polyhedra from Tessellations in Curved Space}%

\author{Philipp W. A. Sch\"onh\"ofer$^1$, Kai Sun$^2$, Xiaoming Mao$^2$, and Sharon C. Glotzer$^{1,2,3}$}%
\email{sglotzer@umich.edu}
\affiliation{Department of Chemical Engineering, University of Michigan, Ann Arbor, Michigan 48109, USA.}
\affiliation{Department of Physics, University of Michigan, Ann Arbor, Michigan 48109, USA.}
\affiliation{Biointerfaces Institute, University of Michigan, Ann Arbor, Michigan 48109, USA.}


\begin{abstract}
\normalsize
Entropic self-assembly is governed by the shape of the constituent particles, yet \textit{a priori} prediction of crystal structures from particle shape alone is non-trivial for anything but the simplest of space-filling shapes. At the same time, most polyhedra are not space filling due to geometric constraints, but these constraints can be relaxed or even eliminated by sufficiently curving space. We show using Monte Carlo simulations that the majority of hard Platonic shapes self-assemble entropically into space-filling crystals when constrained to the surface volume of a 3-sphere. As we gradually decrease curvature to ``flatten'' space and compare the local morphologies of crystals assembling in curved and flat space, we show that the Euclidean assemblies can be categorized either as remnants of tessellations in curved space (tetrahedra and dodecahedra) or non-tessellation-based assemblies caused by large-scale geometric frustration (octahedra and icosahedra).
\end{abstract}

\maketitle

\textit{Introduction} -- Particle shape has become an important design parameter in material science \cite{CKM2007,LKKLJK2019,WRMWL2018}, colloidal self-assembly \cite{JMLZYSM2010,ZLYvdLG2013,UPWRLM2021} and granular matter \cite{ZM2008,WSBRSSB2014,MDJ2019}. One example of the importance of shape is systems of hard particles, which, due solely to entropy maximization, can self-assemble into a zoo of different colloidal crystals simply by changing, even subtly, particle shape \cite{DEG2012}. Although methods exist to inversely design particle shapes likely to self-assemble into targeted crystalline structures \cite{vAKKDG2015,GvADDG2019,MKdPJ2016,CBFD2022}, it is non-trivial to predict those structures from particle shape, other than through molecular simulation. This challenge becomes apparent even for the simplest polyhedra, the Platonic solids. Particle shape directly determines both their assemblies \cite{DEG2012,VG2022} -- in which entropy is maximized -- and \textit{packings} \cite{DEG2012} -- in which density is maximized. However, assemblies and packings are the same only for some Platonic solids and polyhedra in general \cite{SDEG2015}. Cubes are one example where the self-assembled simple cubic (SC) crystal structure and densest packing (a space-filling SC crystal) coincide \cite{SFMD2012}. Hard octahedra and icosahedra do not fill 3D space, but they, too, maximize entropy in crystals that coincide with their densest packing structures: a rhombohedral and face-centered cubic (FCC) structure, respectively \cite{TJ2009,DEG2012}. However, hard dodecahedra self-assemble into a 20-particle unit cell $\beta$-Manganese rotator crystal \cite{DEG2012} with two distinct local particle environments instead of its densest packing structure, FCC \cite{TJ2009}. Likewise, hard tetrahedra famously form quasicrystals \cite{H-AEG2011,H-AEG2011_2} with a myriad of different particle environments instead of the putative densest packing structure with a unit cell comprised of four tetrahedra arranged in a double-dimer structure \cite{CEG2010, H-AEG2011}. Evidently, densest packings, at least in Euclidean space, cannot serve as indicators to predict self-assembly \cite{CKEDG2014,KCEG2018}. But what of curved space?
\begin{figure*}[t!]
\label{fig:polytopes}
\centering
\includegraphics{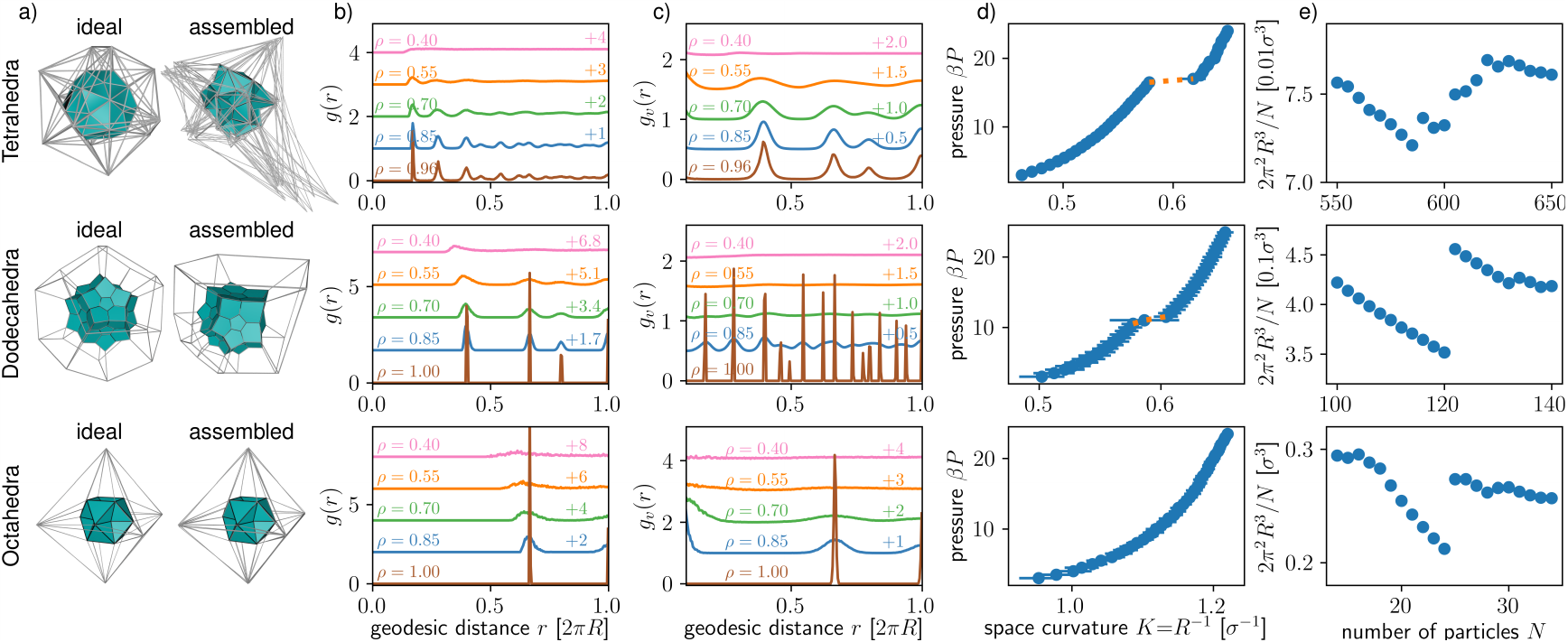}
\vspace{-0.3cm}
\caption{\small Self-assembly of 600 hard tetrahedra into the 600-cell (top row), 120 hard dodecahedra into the 120-cell (center row) and 24 hard octahedra into the 24-cell (bottom row) on the 3-sphere. a) Stereographic projections of the ideal 4-polytopes and the densest obtained assembled configuration via MC simulations. Particles that are highly deformed by the stereographic projection are outlined by their edges for better visualization. Normalized radial distribution functions at different densities $\rho$ in regard to the b) center positions and c) vertex positions of the particles. d) Space curvature vs pressure calculations during the phase transition. e) Highest densities obtained from self-assembly simulations at different number of particles $N$.}
\vspace{-0.6cm}
\end{figure*}

Polyhedra fail to self-assemble their densest packing structures when they fail to resolve global geometric constraints that prevent the polyhedra from maximizing entropy locally as well as globally \cite{vAKAEG2014}. For example, entropy is maximized for cubes when cube faces are aligned, a motif consistent with the SC densest packing, and thus no geometric constraints arise during assembly. Inspired by studies of Frank-Kasper phases \cite{K1989,T2017}, glasses\cite{N1983,N1983_2,MS1984,VS1985,S1984}, tetrahelix sheets \cite{SLKSM2021} and liquid crystal blue phases \cite{SWM1983,S1985,CAAKS2013,SMS2020}, we investigate in this manuscript if and how hard particle assemblies are related to space filling tessellations of curved space. We hypothesize that, if we can find a suitable space with curvature $K$ that permits a shape to tessellate, then the shape will self-assemble into a crystal based on the tessellation in that space because the tessellating arrangement will maximize entropy. By subsequently flattening the space and monitoring the defects that arise in the process, we posit that we will gain predictive information on the likely structure of the Euclidean (3D) assembly. 

We first tested our hypothesis by determining if, in positively curved space, any of the five Platonic solids self-assemble entropy maximizing, tessellating 4-polytopes with no global geometric frustration. We performed hard particle Monte Carlo (HPMC) simulations and show that tetrahedra, dodecahedra and octahedra self-assemble into their corresponding 4-polytopes, each in a differently curved space. We then geometrically frustrated the assemblies by simultaneously increasing the 3-sphere radius $R{=}K^{-1}$ while keeping packing fraction constant, thereby flattening the curved spaces. By comparing the local environments of particles assembled in curved and in flat space, we show that the geometric incommensurability, that prevents particles to form entropically favorable tessellations, manifests itself in two different ways as we gradually flatten space. Interestingly, the Euclidean assemblies of tetrahedra and of dodecahedra still exhibit signs of their 3-sphere tessellations. The geometric frustration suffered by the tessellating assembly as curved space is flattened resolves by the appearance of defects, leading to an assembly with free volume distributed non-uniformly through the structure. In contrast, we find that the assemblies of octahedra and of icosahedra have a large curvature mismatch between the curved spaces they tessellate and Euclidean space, and thus their 3D assemblies are not related to defect-ridden tessellations from curved space. Instead, these shapes resolve the geometric frustration in Euclidean space by maximizing entropy uniformly among all the particles, resulting in colloidal crystals of considerably less complexity than those assembled by tetrahedra and by dodecahedra.\\

\textit{Self-Assembly of 4-polytopes} -- The family of regular 4-polytopes can be identified as tessellations of the 3D positively curved volume of a 3-sphere. We performed HPMC simulations of the self-assembly of $N{=}600$ tetrahedra, $N{=}120$ octahedra and $N{=}24$ dodecahedra with circumsphere diameter $\sigma$ confined to the 3-sphere into 4-polytopes corresponding to the 600-cell consisting of 600 tetrahedral cells, the 120-cell with 120 dodecahedral cells and the 24-cell with 24 octahedral cells, respectively. There exist even more tessellations of the three Platonic solids both in hyperbolic and spherical space, but these three tessellations deviate in curvature the least from flat Euclidean space, which makes them the strongest candidates for a comparison with self-assembled 3D structures. Because cubes already tessellate Euclidean space, and icosahedral tessellations exist only in hyperbolic space, we do not simulate these two shapes \footnote{A first attempt to perform simulations in hyperbolic space \cite{MK2007} investigated only hard spheres.}.
All self-assembly simulations were carried out at constant pressure and constant $N$. Fig.~1d shows equations of state for all three shapes. The data indicates a first-order transition from the disordered fluid phase into a crystalline phase for the tetrahedron and dodecahedron systems. Although a first-order transition is not evident in the octahedron data, we suspect this is simply due to the necessarily small system size \cite{BL1984,PE1988}.

The crystal structures that self-assemble above the transition pressure (or corresponding density) are quantified by two different types of radial distribution functions (RDF; see Fig.~1b and c). The RDF $g_c(r)$ quantifies spatial correlations between particle centroids, and develops peaks that coincide with the characteristic geodesic distances between cells of the ideal 4-polytopes. Similarly, the RDF $g_v(r)$ calculated from the polyhedron vertices develops peaks that fit the dual lattices of the 600-cell (dual: 120-cell), 120-cell (dual: 600-cell) and 24-cell (self-dual). Moreover, we observe that during the formation of the 120-cell, the dodecahedron particles first achieve translational order before they align their orientations, indicating a transition from the isotropic phase into a plastic 120-cell and then into a 120-cell crystal.

By further increasing the density in the 24 octahedra and 120 dodecahedra system, the peaks of $g_c(r)$ and $g_v(r)$ narrow into delta functions, indicating space-filling ideal packings. Also, the stereographic projections of the self-assembled structures reveal the formation of the tessellations (see Fig.~1a and Movies 1-3). We were unable to compress the 600 tetrahedra into the perfect tiling of the 3-sphere, instead, we observe a 600-cell with void defects and interstitials at a maximum density $\rho{=}0.96$.\\
\begin{figure}[t!]
\centering
\includegraphics[width=0.5\textwidth]{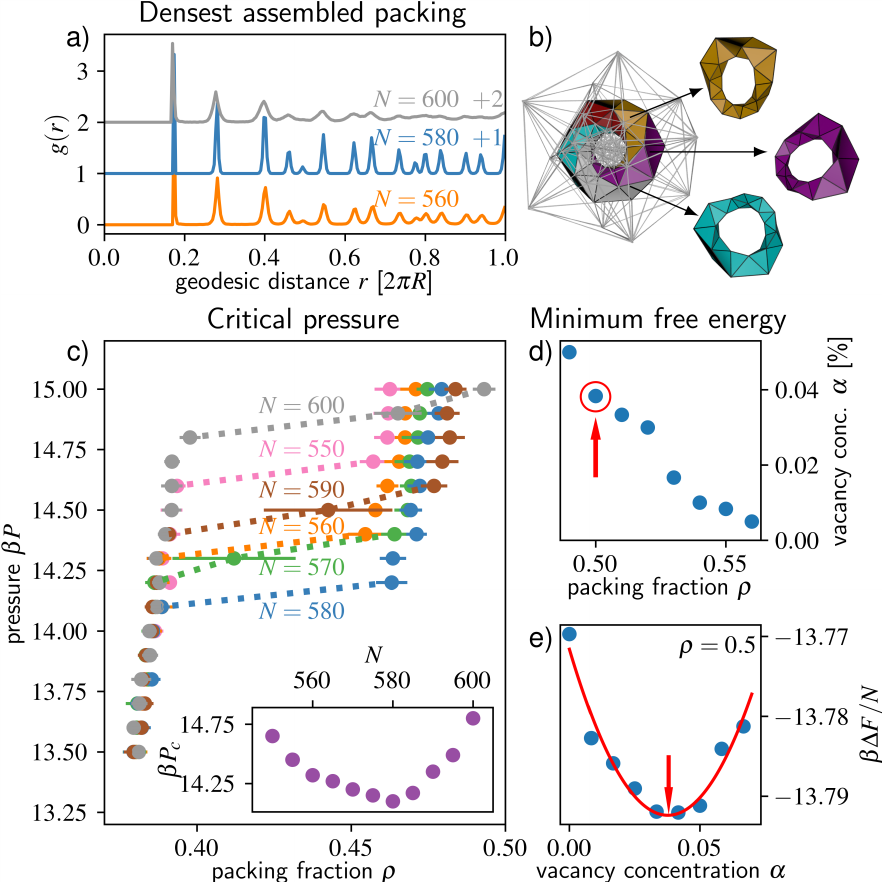}
\vspace{-0.8cm}
\label{fig:Free_energy}
\caption{\small a) Normalized radial distribution functions at the highest obtained density for 560, 580 and 600 hard tetrahedra on the 3-sphere. b) Stereographic projections of tetrahelix loops extracted from an ideal 600-cell. c) Pressure calculations during the phase transition for different numbers of hard tetrahedra on the 3-sphere. The inset plot shows the critical densities at the phase transition. d) Vacancy concentration with the lowest per-particle free energy at different packing fractions. We used the Frenkel-Ladd method (see SI) to calculate the free energy difference $\Delta F$ relative to an Einstein crystal. e) Free energy difference at different vacancy concentrations for $\rho{=}0.5$. The red arrows indicate the relation between d) and e).}
\vspace{-0.6cm}
\end{figure}

\textit{Defect stabilized 600-cell} -- To determine why the perfect 600-cell tessellation of tetrahedra does not assemble at high densities we performed additional simulations with slightly lower and higher numbers of particles (see Fig.~1e). Whereas the octahedron and dodecahedron 3-sphere systems are the most densely packed for the ideal number of particles that correspond to their 4-polytopes, the tetrahedron systems create the lowest local density for $N{=}585$ when defects are present (see Movie 4). Also the critical pressures at the phase transition and free energy calculations in Fig.~2 indicate that the 600-cell spherical lattice is stabilized by impurities at the transition with $N{=}585$. This stabilization of the crystalline phase via vacancies is similar to the equilibrium SC phase of hard cubes \cite{SFMD2012}, where cubes form linear arrays that can slide along each other adding another entropic contribution and leading to the stabilization of the SC crystal via the inclusion of void defects. Analogously, the 600-cell can be separated into 20 linear arrays known as tetrahelix loops, indicating a similar sliding mechanism \cite{S2001} (see Fig.~2b). Despite our free energy calculations, which suggest that the hard tetrahedron system should eliminate the vacancies at higher densities like hard cubes do in Euclidean space, the hard tetrahedron system confined to the 3-sphere is configurationally trapped. Our compression scheme without allowing temporary overlaps \cite{H-AEKZPP-MG2009}, therefore, is unable to eliminate defects integrated into the crystal structure after its initial assembly, resulting in a defected 600-cell.\\
\begin{figure}[t!]
\centering
\includegraphics[width=0.5\textwidth]{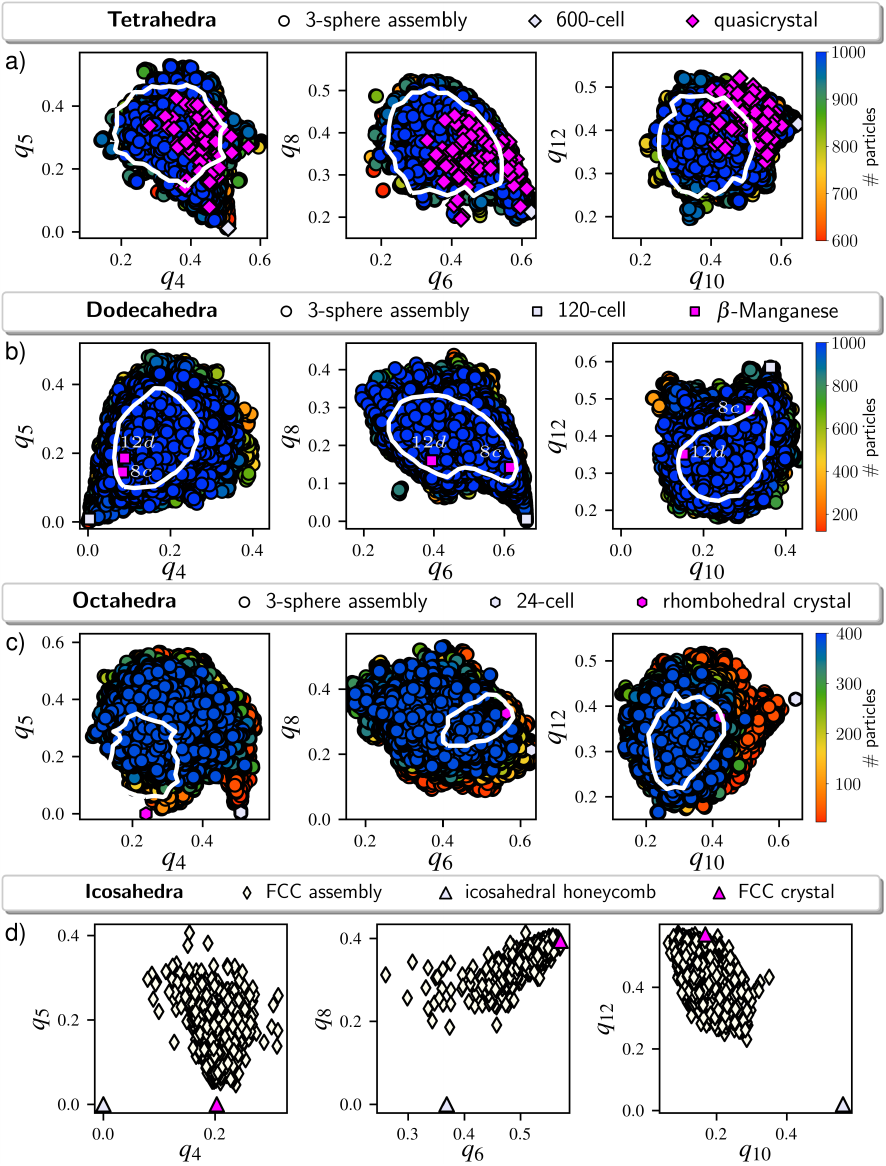}
\vspace{-0.8cm}
\label{fig:Steinhardt}
\caption{\small Minkowski order parameter of hard a) dodecahedron, b) tetrahedron and c) octahedron assemblies on the 3-sphere at $\rho{=}0.65$ when flattening into Euclidean space. Each circle represents a particle environment obtained from the simulation and is compared to crystal structures in Euclidean space. The white outline indicates the region within which lie 80\% of the typical local particle environments of the self-assembled structures in Euclidean space ($\beta$-manganese for dodecahedra with Wyckoff sites $8c$ and $12d$, quasicrystal for tetrahedra and rhombohedral crystals for octahedra). For the tetrahedra we use the (3, 4, 3$^2$, 4) quasicrystal approximant of the dodecagonal quasicrystal as a reference as both structures feature equivalent local environments \cite{H-AEG2011}. d) Comparison between the Minkowski order parameter of hard icosahedra in their densest packings (flat space: FCC, hyperbolic space: icosahedral honeycomb) and a self-assembled FCC structure in Euclidean space.}
\vspace{-0.6cm}
\end{figure}

\textit{Bending into flat space} -- To study how assemblies in Euclidean space resolve geometrical incompatibilities so that particles can arrange into entropically favored configurations, we frustrated the 24-, 120- and 600-cell structures from curved space into Euclidean space. Specifically, we increased the number of particles on the 3-sphere, which simultaneously decreases the space curvature $K{=}\left(\frac{2\pi^2\rho_N}{N}\right)^{\frac{2}{3}}$ at a constant number density $\rho_N$. As the flattening systems gradually incorporate more particles, the larger the 3-sphere radius deviates from the ideal curvature that allows for tessellation. We repeated all simulations with $N_\text{oct}{\in}[26,480]$, $N_\text{dod}{\in}[125,1000]$ and $N_\text{tet}{\in}[620,1000]$. We quantified the assemblies locally by calculating a set of Minkowski-weighted Steinhardt order parameters (SOP) $q_4$, $q_5$, $q_6$, $q_8$, $q_{10}$ and $q_{12}$ \cite{SNR1983,MKS-TM2013} in Fig.~3 (see SI). If the number of polyhedra fits the number that can tessellate the 3-sphere perfectly, the particles achieve a local environment that is in accordance with the 4-polytope configuration once they enter the ordered state. When we slightly increase the number of particles in all three systems, local environments are introduced that deviate from their ideal 4-polytope arrangements. Consequently, the region of typical particle environments expands in the 6-dimensional SOP space, while most particles keep 4-polytope-like environments (see Fig.~3).

\begin{figure}[t!]
\centering
\includegraphics[width=0.5\textwidth]{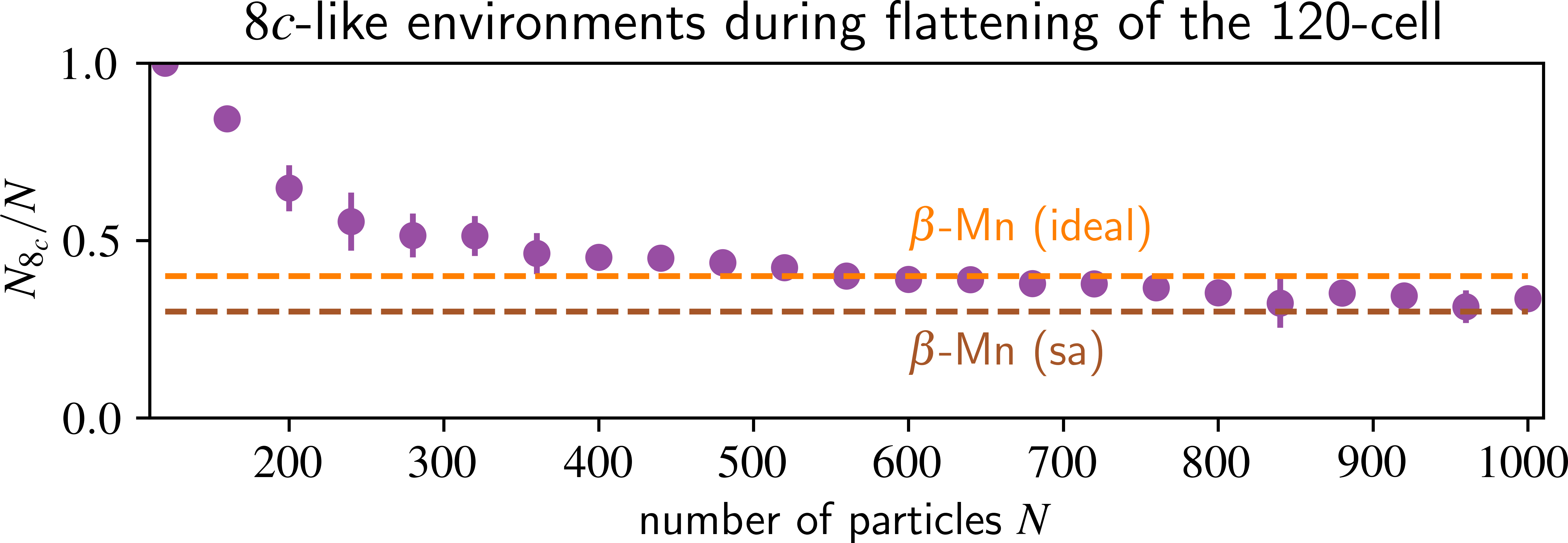}
\vspace{-0.8cm}
\label{fig:ClusterRatio}
\caption{\small Ratio of hard dodecahedra in assemblies on the 3-sphere with local environments closer to the $8c$ rather than the $12d$ Wyckoff site of the ideal $\beta$-Manganese structure in SOP space. The same data is used as in Fig.~3b. The orange dotted line refers to the ideal ratio of $8c$ particles in the $\beta$-Manganese crystal $\left(\frac{N_{8c}}{N}\right)_\text{ideal}{=}0.4$. The red dashed line refers to the obtained ratio when we apply the same calculations to a self-assembled $\beta$-Manganese structure of hard dodecahedra in Euclidean space $\left(\frac{N_{8c}}{N}\right)_\text{sa}{=}0.31{\pm}0.02$.}
\vspace{-0.6cm}
\end{figure}
For the dodecahedron and tetrahedron systems, this region of local environments characterized by SOPs remains consistent even for a large number of added particles (i.e. considerable flattening) and can also be identified as the typical environments of their representative ideal and thermalized self-assembled structures in Euclidean space: the $\beta$-Manganese structure for dodecahedra and the quasicrystal for tetrahedra. Moreover, the development of the local environments with decreasing $K$ indicates why these systems feature multiple local particle arrangements in Euclidean space. The unit cell of the $\beta$-Manganese crystal lattice, for example, contains 20 atoms with two unique Wyckoff sites $8c$ and $12d$ and, hence, two different local environments. The Wyckoff sites are located in two different regions within the SOP space, where, remarkably, $8c$ is close to the ideal 120-cell environment for hard dodecahedra. By assigning each environment during the flattening process to one of the Wyckoff positions depending on their distance in SOP space (see Fig.~4), we identify $8c$ as inherited from the 3-sphere tessellation whereas $12d$ is a disclination integrated into the 120-cell structure. In assemblies with a small number of added dodecahedra $N{\in}[120,200]$ the particles arrange mostly in an $8c$-like local environment with only a few particles accumulating around the $12d$ environment. By flattening space further, more defects arise, which is in accordance with the increase in $12d$-like environments. The ratio between $8c$-like particles converges towards a value between the ideal ratio of sites in a $\beta$-Manganese crystal $\frac{N_{8c}}{N}{=}\frac{8}{20}{=}0.4$ and a ratio obtained from a self-assembled $\beta$-Manganese crystal $\frac{N_{8c}}{N}{=}0.31{\pm}0.02$.

Similarly the multiple environments in the quasicrystal of hard tetrahedra can be interpreted as defects. Even by comparing the SOPs between a quasicrystal and the self-assembled 600-cell structure with void defects we detect that the local environments match (see Fig.~SI1). By flattening space, more and more particles obtain the quasicrystalline environments. We, therefore, argue that the quasicrystal is a result of the entropic gain to integrate different defects into the 600-cell that allows for the development of a variety of different local environments that we interpret as vacancies and interstitials in the 600-cell.

However, hard octahedron systems with many added particles $N_\text{total}{>}50$ paint a different picture. Here, the octahedron assembly must overcome a larger curvature difference to flatten the 24-cell ($\Delta K_{24}{\approx}1.571\sigma^{-1} $) compared to the 120-cell ($\Delta K_{120}{\approx}0.776\sigma^{-1} $) or 600-cell ($\Delta K_{600}{\approx}0.764\sigma^{-1}$). Therefore, the strategy of entropy compartmentalization\cite{MAG2021, LVG2023} by adding defects to the 24-cell eventually becomes less efficient than maximizing entropy by forming crystals with only one type of local environment reminiscent of the rhombohedral crystal.
This phenomenon draws similarities to frustration escape in geometrically frustrated assemblies (GFAs) of deformable particles with open boundaries\cite{HG2017,HG2021,SHG2022,HSG2023}. Despite not being dominated by entropy but instead energetic contributions such as particle deformation, binding between building blocks, and boundary energy terms, GFAs also feature an incompatibility between the locally preferred order and global constraints. Similar to the space curvature in our hard particle assemblies, the particle shape rigidity in GFAs dictates if it is energetically more favorable for the system to accumulate stresses without loosing locally preferred order (rigid self-limiting regime) or to escape the frustration by deforming the particles (soft bulk regime). Hence, the self-assembled structure of hard octahedra is not based on a 3-sphere tessellation. Instead the assembly is more related to the densest packing in Euclidean space, which indicates how to accommodate global and local geometric frustration uniformly.

Although we do not perform a similar computational study with icosahedra, we observe in Fig.~3d that the typical local environments of the self-assembled FCC crystal of icosahedra show the same characteristics as the octahedron system. Within the SOP space the icosahedra sit in-between an ideal FCC and the ideal neighborhood of the icosahedral honeycomb (IH) tessellation of hyperbolic 3-space, but are clearly detached from the latter. This suggests that, like the octahedron system, the occurrence of the FCC crystal in hard icosahedron systems is also caused by the large difference in curvature between the IH and Euclidean space. As in the octahedron system, the frustration of being geometrically restricted from tessellating Euclidean space is distributed uniformly. Therefore, we surmise that the FCC phase is not related to the IH tessellation in hyperbolic space, but rather to the densest packing in Euclidean space.\\

\textit{Conclusion} -- In this Letter, we related the assembly of complex colloidal crystal structures of hard polyhedra to tessellations in curved space. By performing MC simulations of hard tetrahedra, octahedra and dodecahedra on positively curved 3-spheres, we showed that the particles thermodynamically self-assemble their 4-polytope tessellations. This observation indicates that the particles will adopt their locally optimal configurations based on tessellations if no geometrical restrictions are present, such as those that occur for these systems when they attempt to crystallize in Euclidean space. Moreover, we observed by flattening space that the equilibrium colloidal crystal structures of the Platonic solids in Euclidean space can be separated into two categories. The first category includes entropy maximizing, self-assembled structures in flat space that can be understood as remnants of perfect tessellations in curved space. The dodecagonal quasicrystal reported in hard tetrahedron assemblies can be traced back to the 600-cell and its low entropic cost to implement defects into the crystal. Likewise, the $\beta$-Manganese configuration of hard dodecahedra stems from the 120-cell, with one Wyckoff site of the $\beta$-Manganese crystal identified as a local environment native to the 120-cell tessellation, and the other Wyckoff site identified as a defect of the ideal 120-cell structure. The second category includes self-assembled crystals in 3D space that do not stem from tessellations in curved space, such as the rhombohedral crystal of hard octahedra or the FCC crystal of hard icosahedra. Their corresponding space-filling tessellations require spaces with considerably larger curvature than the tessellations of the tetrahedron or dodecahedron. Consequently, maximizing entropy by introducing defects is only efficient when the system is slightly flattened. In Euclidean space, these assemblies instead adopt a geometric compromise where entropy maximization is achieved uniformly through a single type of local environment. Although we focused only on hard shapes so far, our findings probably also apply to systems with enthalpic contributions considering the mathematical description of Frank-Kapser phases as disclinated 600-cells \cite{K1989,T2017}. Hence, introducing additional degrees of freedom that help resolving geometric frustration -- such as shape deformability or flexible particle bonds -- might allow us to broaden the curvature window, where assemblies based on non-Euclidean crystals minimize free energy, and guide the prediction of new self-assembly structures.\\

\section*{Acknowledgements}

The authors thank Nicholas Kotov, Nan Cheng and Francesco Serafin for helpful discussions. X.M. and K.S. were supported in part by the Office of Naval Research (MURI N00014-20-1-2479), and by the National Science Foundation (NSF PHY-1748958); P.S. and S.G. were supported by a grant from the Simons Foundation (256297, SCG). This work used the Extreme Science and Engineering Discovery Environment (XSEDE), which is supported by National Science Foundation grant number ACI-1548562; XSEDE award DMR 140129. Computational resources and services were also supported by Advanced Research Computing at the University of Michigan, Ann Arbor.

\bibliographystyle{unsrt.bst}
\bibliography{reference.bib}
\end{document}


\title{Supplementary information:\\Rationalizing Euclidean Assemblies of Hard Polyhedra from Tessellations in Curved Space}%

\author{Philipp W. A. Sch\"onh\"ofer$^1$, Kai Sun$^2$, Xiaoming Mao$^2$, and Sharon C. Glotzer$^{1,2,3}$}%
\email{sglotzer@umich.edu}
\affiliation{Department of Chemical Engineering, University of Michigan, Ann Arbor, Michigan 48109, USA.}
\affiliation{Department of Physics, University of Michigan, Ann Arbor, Michigan 48109, USA.}
\affiliation{Biointerfaces Institute, University of Michigan, Ann Arbor, Michigan 48109, USA.}

\maketitle

\section{Methods}
We perform Monte Carlo simulations of hard polyhedra with circumsphere diameter $\sigma$ in positively curved three-dimensional space. We model the 3-dimensional space with constant curvature $K$ as a spherical confinement in four dimensions such that all points $\mathbf{r} = (w,x,y,z)$ of the spherical polyhedra are embedded in the surface volume of a hypersphere with radius $R = \frac{1}{K}$
\begin{equation}
\label{eq:EqMotionT}
R^2 = w^2 + x^2 + y^2 + z^2.
\end{equation}
Computationally, these simulations are realized by implementing hyperspherical boundary conditions into Hoomd-blue v2.9 \cite{AGS2020} and representing both the position and orientation of each polyhedron by a pair of quaternions. The algorithm allows for three Monte Carlo moves: translation via parallel transportation along geodesic great circle lines, local rotation around a randomly chosen axis and reflections at the face of the polyhedron. The exact scheme is detailed in Ref. \cite{SBL2012}.

A configurational change within a Monte Carlo step is accepted based on the excluded volume between particles. Hence, to determine the overlap between two polyhedra we considered geodesic distances on the hypersphere surface volume instead of 4-dimensional Euclidean distances. This means that the whole bodies of the polyhedra live on the hypersphere surface. Consequently, the edges and faces spanned by the vertices all lie on geodesic lines and geodesic planes, respectively, and are curved according to $R$. Numerically, we check for intersections between particles using a XenoColloide algorithm scheme \cite{S2008} modified to positively curved spaces by applying spherical instead of Euclidean trigonometry. The maximum translational and rotational step is chosen such that around 50\% of change attempts are accepted.

For our simulations we perform multiple sets of $NVT$-simulations. Starting from a set number of particles $N$ we start the simulations at a high hypersphere radius $R=N^{\frac{1}{3}}$ such that the polyhedra are in an isotropic gas phase. After every 1000th Monte Carlo step we attempt to decrease $R$ by a factor of $0.999$ to both slowly increase the packing fraction and the curvature of the space. The attempt is successful if no overlaps have been detected and the simulation continues with the new radius. Otherwise, the system is reset to the radius before the attempt. At every increment of $\Delta R = 0.2\sigma$ the system is equilibrated for $1\times10^6$ Monte Carlo steps to avoid frustration during the process. These compression steps are repeated until 100 consecutive shrinking attempts have failed. For each number of particles, we ran 5 replica simulations.

Additional to the $NVT$-simulations we also run sets of isobaric simulations to determine the phase transition of polyhedra from the isotropic gas phase into ordered crystal structures. For the volume move attempts we increase or decrease the radius of the hypersphere. We choose the maximum radius move size such that roughly 50\% of volume move attempts are successful. We start from a low pressure $P=1$ and increase the pressure by $\Delta P = 0.5$ every $2\times10^6$ steps. For the last $1\times10^6$ Monte Carlo steps on each pressure level we determine the hypersphere radius to determine the phase transitions.

\section{Free energy calculations}
\label{app:FrenkelLadd}
To calculate the per particle free energies $\frac{\beta F}{N}$ of the 600-cell with defects we use the Frenkel-Ladd method \cite{FL1984, FS2001}:
\begin{equation}
\frac{\beta F}{N} = \frac{\beta F^\text{Ein}(\lambda_m)}{N} + \frac{\beta \Delta F}{N}
\end{equation}

We compare the different systems to a non-interacting Einstein crystal $\frac{\beta F^\text{Ein}(\lambda_m)}{N} $ where each particle is harmonically bonded to a site of the ideal 600 cell both in terms of position and orientation with a high coupling strength $\lambda_m = 10000\epsilon$. The free energy difference
\begin{equation}
\frac{\beta \Delta F}{N} = -\frac{1}{N}\log\left(\frac{N_L!}{(N_L-N)!N!}\right) - \frac{\beta}{N}\int_0^{\lambda_m}\Biggl \langle \frac{\partial U^\text{ext}(\lambda)}{\partial \lambda} \text{d}\lambda \biggr \rangle
\end{equation}
consists of a combinatorial first term that takes all possible positions of the vacancies into account and the second external term. We calculate the external term via numerical integration of the harmonic bond potentials 
\begin{equation}
\beta U^\text{ext}(\lambda) = \lambda \sum_{i=0}^N \left(\frac{1}{\sigma^2}|\mathbf{x}_i-\mathbf{r}_0|^2 + (1-\mathbf{u}_i\cdot\mathbf q(\mathbf{r}_0))^2\right)
\end{equation}
where $\mathbf{x}_i$ and $\mathbf{u}_i$ are the position and orientation quaternion of the $i$-th particle and $\mathbf{r}_0$ and $\mathbf{q}(\mathbf{r}_0)$ are the position and orientation quaternion of the closest site of the ideal 600-cell structure. For  $\mathbf{q}(\mathbf{r}_0)$ we considered the symmetries of the particle shape. 

\section{Steinhardt order prameter}
\label{app:Steinhardt}
To determine the local environments within the polyhedra system we calculate the rotationally invariant Minkowski weighted Steinhardt order parameters $q_l$ with $l\in\{3,4,5,6,10,12\}$ for each particle $i$ \cite{SNR1983,MKS-TM2013}. 
\begin{equation}
q_l(i)=\sqrt{\frac{4\pi}{2l+1}\sum_{m=-l}^{l}|q_{lm}(i)|^2}
\end{equation}
The quantity $q_{lm}$ is comprised of a weighted sum over the spherical harmonics $Y_{lm}$ between particle $i$ and $j$
\begin{equation}
q_{lm}(i)=\frac{1}{N_b}\sum_{j=1}^{N_b}w_{ij}Y_{lm}(\theta_{ij},\phi_{ij})
\end{equation}
with the number of neighbors $N_b$, weights $w_{ij}$ and the polar angles $\theta_{ij}$ and $\phi_{ij}$ between the bond of $i$ and $j$.
To calculate the polar angles from the 4-dimensional position data of the particles we map all neighbor particles $j$ to the flat 3-dimensional space that is tangential to the hypersphere and touches it at the position of particle $i$. Afterwards, we determine the angles according to the bond vector $\mathbf{r}_{ij}$ and the orientation of particle $i$. The polar angles are the same on the hypersphere and the tangential space as the mapping is conformal and preserves all angles between directed vectors. 
For Minkowski weighted Steinhardt order parameter the weights are defined as $w_{ij}=\frac{A_{ij}}{A_i}$, where $A_i$ is the surface area of the Voronoi polyhedra around particle $i$ and $A_{ij}$ is the area of the face between particle $i$ and $j$.
We construct the Voronoi tessellation on the hypersphere by calculating the 4-dimensional convex hull around the central position of each particle. As the particles live on the surface volume of the hypersphere, the 4D convex hull is equivalent to a Delauney tetrahedralization in spherical space. The dual lattice of the Delauney tetrahedralisation then corresponds to the Voronoi tessellation. \\

\newpage
\section{Comparison between assembled 600-cell and quasicrystal environments}

\begin{figure}[h!]
\centering
\includegraphics[width=0.8\textwidth]{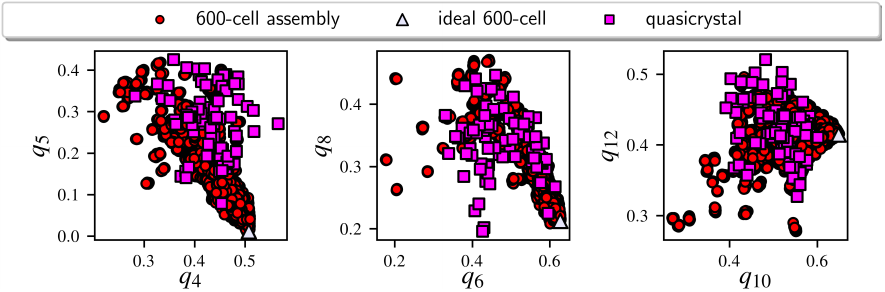}
\caption{\small Minkowski order parameter of hard tetrahedron assemblies on the 3-sphere at the highest density $\rho{\approx}0.96$ obtained from our Monte Carlo simulations. Each circle represents a particle environment obtained from the defected 600-cell (see Fig.~1a in the main document) and is compared to the local environments of the (3, 4, 3$^2$, 4) quasicrystal approximant.}
\vspace{-0.3cm}
\end{figure}

\bibliographystyle{unsrt.bst}
\bibliography{reference.bib}